\documentclass{aastex}
\usepackage{natbib,emulateapj5}
\begin{document}

\def\lapp{\ifmmode\stackrel{<}{_{\sim}}\else$\stackrel{<}{_{\sim}}$\fi}
\def\gapp{\ifmmode\stackrel{>}{_{\sim}}\else$\stackrel{>}{_{\sim}}$\fi}

\title{X-Ray Emission from the Double Pulsar System J0737--3039}
\author{M.\ A.\ McLaughlin\altaffilmark{1},  F.\ Camilo\altaffilmark{2},
M.\ Burgay\altaffilmark{3,5}, N.\ D'Amico\altaffilmark{4,5}, B.\
C.\ Joshi\altaffilmark{6}, M.\ Kramer\altaffilmark{1}, D.\ R.\
Lorimer\altaffilmark{1}, A.\ G.\ Lyne\altaffilmark{1}, R.\ N.\
Manchester\altaffilmark{7} \& A.\ Possenti\altaffilmark{5} }
\altaffiltext{1}{Jodrell Bank Observatory, University of Manchester,
  Macclesfield, Cheshire, SK11 9DL, UK }
\altaffiltext{2}{Columbia Astrophysics Laboratory, Columbia University,
  550 West 120th Street, New York, NY 10027 }
\altaffiltext{3} {Universit\`a degli Studi di Bologna, Dipartimento di
  Astronomia, via Ranzani 1, 40127 Bologna, Italy }
\altaffiltext{4} {Universit\`a degli Studi di Cagliari, Dipartimento di
  Fisica, SP Monserrato-Sestu km 0.7, 09042 Monserrato, Italy }
\altaffiltext{5}{INAF-Osservatorio Astronomico di Cagliari, loc. Poggio
  dei Pini, Strada 54, 09012 Capoterra, Italy }
\altaffiltext{6}{National Centre for Radio Astrophysics, P.O. Bag 3,
  Ganeshkhind, Pune 411007, India }
\altaffiltext{7} {Australia Telescope National Facility, CSIRO,
  P.O.~Box~76, Epping, NSW~1710, Australia }

\begin{abstract}
We report on a 10\,ksec observation of the double pulsar system
J0737--3039 obtained with the {\em Chandra\/} X-ray Observatory's
Advanced CCD Imaging Spectrometer. We detect the system as a point
source with coordinates (J2000) $\mbox{RA} = 07^{\rm h}37^{\rm
m}51\fs23$, $\mbox{Decl.} = -30\arcdeg39'40\farcs9$, making this the
first double neutron star system to be detected at X-ray energies.
Only 77 source counts are detected.  The data are represented
satisfactorily by a power-law spectrum with photon index $\Gamma =
2.9 \pm 0.4$.  The inferred 0.2--10\,keV luminosity is $\approx
2\times10^{30}$\,erg~s$^{-1}$ for a distance of 0.5\,kpc.
We do not detect any significant variability with orbital phase in these
data, which span one full orbit. These results are consistent with the
X-ray emission originating solely from the magnetosphere of the energetic
A pulsar, with an efficiency for conversion of rotational energy to
X-rays in this band of $\sim$ 0.04\%, although we cannot exclude other possibilities.
\end{abstract}


\section{Introduction} \label{sec:intro}

Thirty years following the discovery of the first double neutron star
\cite{ht75}, only eight such systems are known.  A few of these rare
binaries are wonderful laboratories for the study of relativistic gravity
and gravitational radiation.  One of the more recently discovered of
these systems, J0737--3039 \cite{burgay03}, is fundamentally different
from all of the others observationally in that it contains {\em two\/}
detected pulsars \cite{lyne04}.  The first pulsar discovered in the system
(A) spins every 22\,ms, has a low inferred surface magnetic field strength,
a characteristic age of 210\,Myr, and a relatively high rate of
rotational energy loss, $\dot E = 5.8\times10^{33}$\,erg~s$^{-1}$.
Pulsar B has a period of 2.7\,s, a characteristic age of 50\,Myr,
and an $\dot E$ that is 3600 times smaller than that of A.  Its mass,
1.25\,M$_{\odot}$, is the lowest known for any neutron star and is
0.09\,M$_{\odot}$ less than that of A.  Both pulsars orbit their common
center of mass every 2.4\,hr with eccentricity of 0.09. We observe the
system nearly edge-on, with an  orbital inclination of $87\arcdeg$.

A unique aspect of the system is that radio emission from pulsar B
is only detected strongly in two repeatable orbital phase ranges, each
of duration $\sim 10$\,min.  A $\sim 30$\,s-long eclipse of pulsar A
is also observed at its superior conjunction.  Both of these effects
most likely result from the interaction of the fast-spinning pulsar's
relativistic wind with the magnetosphere of its much less energetic
companion \cite{lyne04}.

This system will be an exceptional laboratory for relativistic
astrophysics and should provide unique insights into the magnetospheres
and the local environment of both neutron stars.  High-energy
observations, particularly at X-ray energies, could be especially valuable
for understanding the energetics of the system.  Whether resulting
from magnetospheric or thermal processes, most pulsars emit of order
$10^{-3}\dot{E}$ in soft X-rays \cite{bt99}. This translates to an
expected luminosity of order $10^{31}$\,erg~s$^{-1}$ for pulsar A,
and means that B should be undetectable due to its very low $\dot
E$ and relatively large age.  But in this particular system, while
still unlikely, it is conceivable that a portion of A's relativistic
Poynting/particle flux might heat the surface of B, perhaps causing it
to emit detectable radiation.  Such emission could prove invaluable for
probing the atmosphere of B, which is likely to deviate significantly
from standard models.  A more likely source of ``unusual'' X-ray
emission might be the shock that undoubtedly forms near pulsar B owing 
to the collision between A's relativistic wind and B's magnetosphere.
Some of this
emission could be time variable, for instance due to Doppler boosting.

Many of these emission modes are clearly speculative and hard to
quantify with the present understanding of the system.  Nevertheless, given
the potential for unusual insights into the physics of this unique
interacting binary, we have obtained a short X-ray observation of this
system, awarded through the {\em Chandra\/} Director's Discretionary
Time program, in order to better plan future multi-wavelength studies.
In this Letter we report the detection of a point source at the position
of the J0737--3039 system and describe its properties.

\clearpage
\section{X-Ray Observations and Analysis} \label{sec:xray}

\begin{figure}
\epsscale{0.6}
\plotone{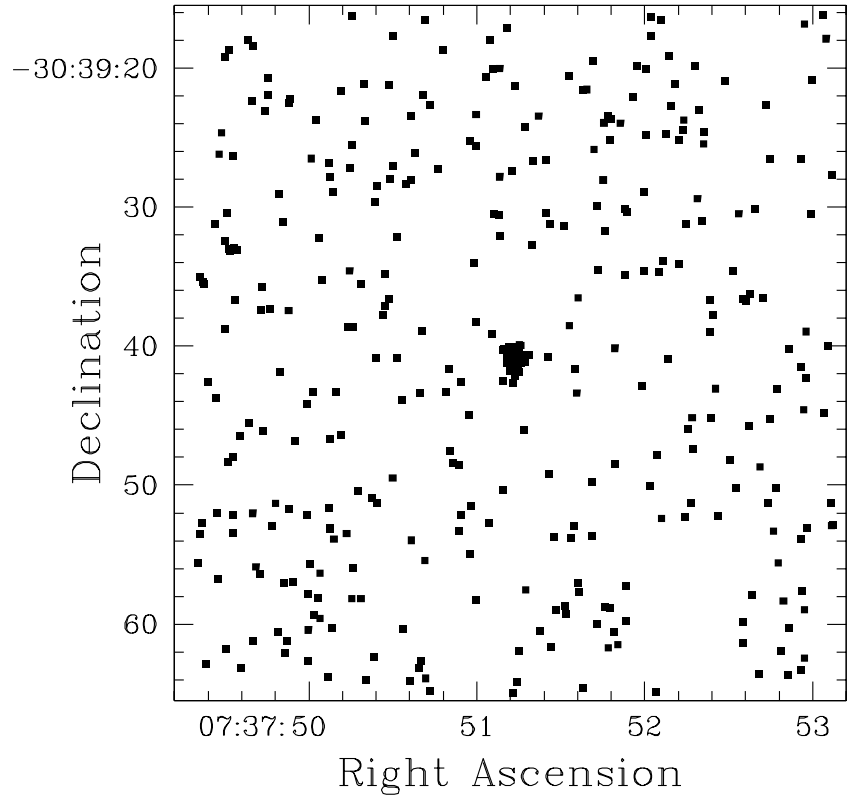}
\caption{
\label{fig:image}
A $50''\times50''$ portion of the ACIS-S3 chip in the 0.2--10\,keV band,
with the double pulsar J0737--3039 at its center.  No other sources were
detected in the entire $16'\times16'$ area of the chip. }
\end{figure}

The double pulsar system J0737--3039 was observed with the Advanced
CCD Imaging Spectrometer S-array (ACIS-S) instrument aboard {\em
Chandra\/} on 2004 January 18 for 10\,ksec (corrected exposure time
was 10,009\,s).  The system's position was located on the aim point of
the ACIS S3 chip.  We used the \verb+CIAO+ package\footnote{Available
at http://asc.harvard.edu/ciao.}, version 3.0.2, to perform much of the
analysis of these data, using the most recent release of the calibration software,
\verb+CALDB 2.26+, which corrects for the loss of quantum efficiency at low energies
due to molecular contamination.  As shown in Figure~\ref{fig:image}, we detect a
source at (J2000) $\mbox{RA} = 07^{\rm h}37^{\rm m}51\fs23$, $\mbox{Decl.}
= -30\arcdeg39'40\farcs9$.  No other sources are detected in the entire
S3 chip.  This position is $0\farcs3$ away from that derived from radio
timing of the pulsars \cite{lyne04}, well within the X-ray $1\,\sigma$
position uncertainty of $0\farcs8$, dominated by the satellite pointing
error of approximately $0\farcs7$.  We have therefore detected the
J0737--3039 system in X-rays.  Using a circular aperture of radius $2''$
centered on the position of the system, and four background regions, we
determine that $77 \pm 9$ counts originate from the pulsar, corresponding
to a rate of $(7.7 \pm 0.9)\times 10^{-3}$ counts/s. The system
is detected as a point source, with the spatial spread consistent with
the point spread function of the instrument.

\begin{figure}
\epsscale{0.6}
\plotone{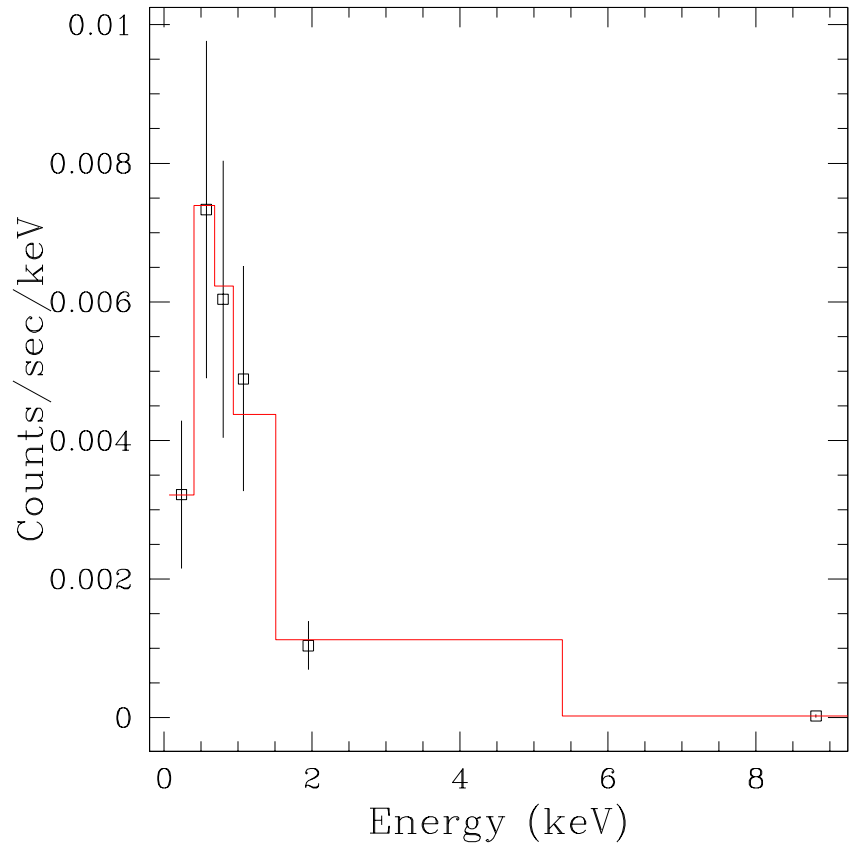}
\caption{
\label{fig:spectrum}
The background-subtracted spectrum of J0737--3039, with the power-law
fit described in \S~\ref{sec:xray}.  Each of the first five energy bins contains
15 counts; the last bin contains two counts.
The squares are placed at the average energy of the photons in each bin,
while the solid best-fit line illustrates each bin's full width. Error
bars, calculated by {\em Sherpa\/}, account for uncertainties in
both source counts and the instrumental model.}
\end{figure}

We used \verb+psextract+ to obtain a spectrum for this source
and fit it using {\em Sherpa\/}, the \verb+CIAO+ modeling
and fitting package.  As shown in Figure~\ref{fig:spectrum}, a steep
power-law spectrum fits the data satisfactorily, with photon index
$\Gamma \approx 2.9$ and neutral hydrogen column density $N_H \approx
5\times10^{20}$\,cm$^{-2}$.  Because
of the small number of counts we use the maxium likelihood method to fit
these data, implemented in {\it Sherpa} with the CSTAT statistic \cite{cash79}.
This fit returns a Q-value (i.e., the probability that we would
observe the measured CSTAT statistic or higher if the best-fit model is true)
of 0.96. The best-fit blackbody and Raymond-Smith
models return Q-values of $3\times10^{-5}$ and $2\times10^{-5}$.
A thermal bremsstrahlung model returns a Q-value of 0.55, but the best-fit value of $N_H \sim
1\times10^{19}$~cm$^{-2}$ is unreasonably low.  Fitting for multi-component
models or models with more free parameters is unreasonable due to the
small number of counts available. While
 there are only two counts
with energies greater than 3\,keV, the fitted parameters do not change substantially if these counts are removed.
Similarly, the fit does not change
substantially if we exclude photons with energies less than 0.5~keV, where ACIS-S calibration is somewhat uncertain. 

\begin{figure}
\epsscale{0.5}
\plotone{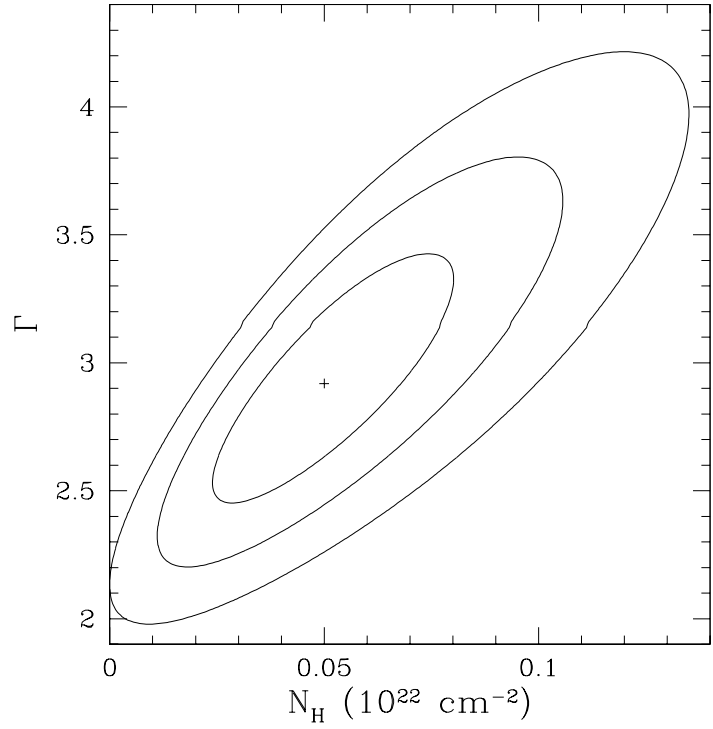}
\caption{
\label{fig:contours}
Contour plot of the CSTAT statistic as a function of the neutral hydrogen column
density $N_H$ and the photon index $\Gamma$. The plotted confidence levels
are 1, 2 and 3$\sigma$.}
\end{figure}

Figure~\ref{fig:contours} shows the a
contour plot of the CSTAT statistic for the best-fit model in the $N_H - \Gamma$ plane. 
 From the derived count rate and
spectral fit, we calculate a 0.2--10\,keV unabsorbed flux of $f_x
\approx 8\times10^{-14}$\,erg~cm$^{-2}$~s$^{-1}$. 
Assuming isotropic emission, and for a distance $d \sim0.5$\,kpc, inferred from its 
dispersion measure (i.e. free electron
column density) and a model for the Galactic electron density \cite{cl02},
we  obtain a 0.2--10\,keV luminosity
of $L_x \approx 2\times10^{30}(d/0.5\,\mbox{kpc})^2$\,erg~s$^{-1}$.
In Table~\ref{tab:xrayprops} we list all relevant spectral quantities
and their uncertainties.  We stress that $d$, and hence the derived
luminosity, may be uncertain by a factor of two or more due to unmodeled
electron density features associated with its position near the Gum
Nebula \cite{mr01}.  VLBI observations will be important for determining
the actual distance.

The value of $N_H$ derived for the system, while rather uncertain, is much smaller 
than the total Galactic value of $4.65\times10^{21}$~cm$^{-2}$ derived by Dickey \& Lockman (1990)
from neutral hydrogen measurements. It is also smaller (at $> 1\sigma$) 
than the value of $1.5\times10^{21}$~cm$^{-2}$
derived from its dispersion measure, assuming 10 H atoms for each $e^-$. Our derived $N_H$,
however, is marginally greater 
than expected given the measured neutral hydrogen
absorption to other sources with similar longitude, latitude and $\sim
0.5$\,kpc distance\footnote{See the ISM Column Density Search Tool at
http://archive.stsci.edu/euve/ism/ismform.html.}, and is more consistent
with the measurements if a distance of at least $\sim1$\,kpc is assumed. Such
discrepancies between $N_H$ measurements and dispersion-measure derived distances for pulsars
are not unusual (e.g. Nicastro et al. 2004).
Also, since the source closest to J0737--3039 with a neutral hydrogen absorption
measurement is a radial distance of $\sim 0.1$\,kpc away, and because the errors on our $N_H$
and distance
measurements are large, further observations are necessary to determine if such a
discrepancy exists.

\begin{figure}
\epsscale{0.5}
\plotone{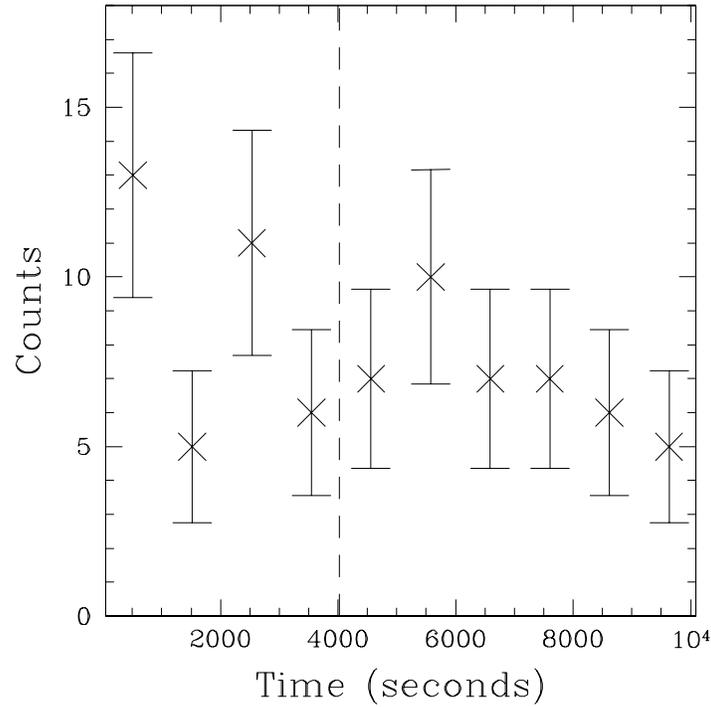}
\caption{
\label{fig:time}
Number of counts (shown with Poisson error bars) vs. elapsed time
(since the start of the observation) with 10 bins across the 10\,ksec
observation.  The orbital period of the system is 8835\,s. Orbital phase
calculated from barycentered data (with respect to the ascending node
of pulsar A) ranged from $281\arcdeg$ at the start of the observation
to $328\arcdeg$ at the end, with the dashed line marking the superior
conjunction of A. }
\end{figure}

In Figure~\ref{fig:time} we plot the number of photons detected in
each of 10 time bins over the 10\,ksec observation (i.e. 1.13 orbital periods).  There is no
significant evidence for variability, with a $\chi^2$ test revealing
only a 51\% probability that the data were drawn from a non-uniform
distribution. We also do not see any correlation between time and the
energy of the photons.  The small number of detected photons makes
searching for variability on shorter time scales impossible.

\section{Discussion} \label{sec:discussion}

We have detected the J0737--3039 system as a faint X-ray source
that is best described
by a steep power-law spectrum with $\Gamma \approx 2.9$ and $L_x
\approx 2\times10^{30}(d/0.5\,\mbox{kpc})^2$\,erg~s$^{-1}$ (see
Table~\ref{tab:xrayprops}).  This result is compatible with the emission
originating solely from the magnetosphere of pulsar A:  Becker \& Tr{\"
u}mper (1999) find that for recycled radio pulsars that are detected at
X-ray energies, $L_x \sim 10^{-3} \dot E$, and that when their spectra can
be fit by a power law, $2 \lapp \Gamma \lapp 2.4$.  While by comparison
our spectral index is a bit steep and 
our efficiency for conversion of rotational to X-ray energies is
a little small ($4\times10^{-4}[d/0.5\,\mbox{kpc}]^2$), uncertainties
in spectral parameters and distance make this interpretation perfectly
plausible.  We note in passing that the X-ray luminosity detected from the
system is roughly equal to the entire spin-down luminosity of pulsar B.

Another intriguing possibility is that these X-rays are produced when
the relativistic winds of A and B collide.  As shown by Arons \& Tavani
(1993), the synchrotron emission produced in such interactions would
also be expected to have a power law X-ray spectrum with $\Gamma \sim 2$.
Unfortunately, the data from this short observation are not sufficient to
discriminate between these two possibilities.  More sensitive observations
with better time resolution are necessary in order to: obtain a better
spectral fit; determine what portion of the detected X-ray flux is
pulsed at the rotation period of pulsar A; quantify better any possible
time variability (the observed radio flux density of B varies dramatically on
$\sim 5$--10\,min timescales that are impossible to investigate with the
present data);  measure any thermal component that may be contributing
to the total X-ray flux of this system.

\acknowledgments

We thank the {\em Chandra\/} X-ray Center for approving the Director's
Discretionary Time for this observation and the anonymous referee for useful comments.
 FC is supported by the NSF
through grant AST-02-05853 and by NASA. MB, NDA and AP received support
from the Italian Ministry of University and Research (MIUR) under the
national program {\em Cofin 2002}.

{}

\begin{deluxetable}{lc}
\tablewidth{0pt}
\tablecaption{X-ray properties of the J0737--3039 system. }
\tablehead{Parameter & Value }
\startdata
$\Gamma$, photon index\dotfill & 
  $2.9^{+0.4}_{-0.4}$ \\
$N_H$, neutral hydrogen column density ($10^{20}$\,cm$^{-2}$)\dotfill &
  $4.8^{+3.4}_{-2.4}$ \\
$f_x$, flux (0.2--10\,keV) ($10^{-14}$\,erg~cm$^{-2}$\,s$^{-1}$)\dotfill & 
  $7.6^{+1.6}_{-1.2}$ \\
$L_x$, luminosity (0.2--10\,keV) ($10^{30}$\,erg~s$^{-1}$)\dotfill &
  $2.3^{+0.5}_{-0.4}$ \\
$L_x/\dot{E}$ (0.2--10\,keV) ($10^{-4}$)\dotfill & 
  $4.0^{+0.8}_{-0.6}$ \\
\enddata
\label{tab:xrayprops}
\tablecomments{Results from a fit to ACIS-S3 data of the form
$dN \propto E^{-\Gamma} dE$. Flux has been corrected for
interstellar absorption. Luminosity assumes a distance of 0.5\,kpc
(see \S~\ref{sec:xray}) and is the same, within the errors, for the 0.2 -- 3~keV energy range.
Uncertainties correspond to $1\,\sigma$
confidence levels.  }
\end{deluxetable}

\end{document}